\documentclass[reprint,amsmath,amssymb,aps,prb]{revtex4-2}
\usepackage[utf8]{inputenc}
\usepackage{dcolumn}
\usepackage{bm}
\usepackage{graphics}
\usepackage[pdftex]{graphicx}
\usepackage[export]{adjustbox}
\usepackage{pgfplots,pgfplotstable}
\usepackage{tikz}
\pgfplotsset{compat=newest}

\begin{document}
\title{A Three-Fluid Model of Dissipation at Surfaces in Superconducting Radiofrequency Cavities}
\author{Michelle M. Kelley}
\author{Sean Deyo}
\author{Nathan Sitaraman}
\author{Danilo B. Liarte}
\author{Tom\'as Arias}
\author{James P. Sethna}
\affiliation{Laboratory of Atomic and Solid State Physics, Cornell University, Ithaca, NY, USA}
\author{Thomas Oseroff}
\author{Matthias Liepe}
\affiliation{Cornell Laboratory for Accelerator-Based Sciences and Education, Cornell University, Ithaca, NY, USA}
\date{\today}

\begin{abstract}
Experiments on superconducting cavities have found that under large RF fields the quality factor can improve with increasing field amplitude, a so-called ``anti-Q slope.'' We numerically solve the Bogoliubov-de Gennes equations at a superconducting surface in a parallel magnetic field, finding at large fields there are surface quasiparticle states with energies below the bulk superconducting gap that emerge and disappear as the field cycles. Modifying the standard two-fluid model, we introduce a ``three''-fluid model where we partition the normal fluid to consider continuum and surface quasiparticle states separately. We compute dissipation in a semi-classical theory of conductivity, where we provide physical estimates of elastic scattering times of Bogoliubov quasiparticles with point-like impurities having potential strengths informed from complementary \textit{ab initio} calculations of impurities in bulk niobium. We show, in this simple yet effective framework, how the relative scattering rates of surface and continuum quasiparticle states can play a role in producing an anti-Q slope while demonstrating how this model naturally includes a mechanism for turning the anti-Q slope on and off.
\end{abstract}

\maketitle

\section{Introduction}
\label{sec:intro}
Superconducting radio-frequency (SRF) cavities are useful in a variety of modern applications, such as free-electron lasers and particle accelerators \cite{Dhakal}. The primary advantage of using superconducting materials over their normal-conducting alternatives is a lower surface resistance, which comes with the benefits of lower operating costs and a lower energy footprint \cite{Padamsee}. Despite decades of steady improvements, leading to quality factors in the neighborhood of $10^{10}$ \cite{Dhakal} and accelerating fields up to $25$--$45$ MV/m \cite{Padamsee}, some open research questions remain. As the accelerating field increases there is often what is referred to as a ``$Q$ slope,'' meaning a quality factor that decreases---or, equivalently, a surface resistance that increases---as the accelerating field increases. However, recently some cavities, in particular niobium cavities doped with nitrogen, have been observed to produce a so-called ``anti-$Q$ slope,'' referring to a quality factor that perplexingly increases with the accelerating field \cite{Dhakal}. Further observations of this phenomenon include stronger anti-$Q$ slopes when increasing the resonant frequency of the cavity \cite{Martinello}, yet, definitive theoretical explanations for the anti-$Q$ slope remain elusive. 

Conventional theories of AC dissipation in superconductors predict quality factors that remain constant as the accelerating field increases, i.e. no $Q$ slope. Mattis and Bardeen~\cite{MattisBar1958} applied linear response methods to BCS theory~\cite{BardeenSch1957} to yield an expression for the complex conductivity of a superconductor subject to a magnetic field. (Here we ignore other sources of dissipation such as trapped magnetic vortices oscillating near the cavity surface~\cite{GurevichCio2013,LiarteSet2018,ChecchinSer2018} and within grain boundaries~\cite{SheikhzadaGur2017,CarlsonPos2021}.)
The result they found was a surface resistance that roughly goes as $\omega^2\exp (-\Delta / k_B T)$ at low temperatures, where $\Delta$ is the superconducting gap and $\omega$ is the angular frequency of the oscillating field. Central to these calculations is the idea that a quasiparticle in a state with energy $E_1$ transitions to a state of energy $E_2=E_1+\hbar\omega$ upon absorbing a photon of energy $\hbar\omega$. In order for the photon energy to be well defined the absorption must happen coherently over many cycles of the AC perturbation.

Extensions of the linear-response theory of~\cite{MattisBar1958} have been routinely implemented in order to interpret experimental data exhibiting anti-$Q$ slopes. Applying the Keldysh formalism in the context of non-equilibrium Green's functions, Gurevich rederived the expression for the  conductivity to evaluate the surface resistance at low frequencies, low mean free paths, and high magnetic fields~\cite{Gurevich}.
There, the anti-$Q$ slope originates from the smearing of the quasiparticle density of states in the presence of an oscillating superflow.
At high enough fields, the superconducting gap decreases and the density of states smears to weaken the zero-frequency singularity from the Mattis-Bardeen expression, which eventually leads to a field dependent surface resistance. This calculation involves an approximate nonequilibrium quasiparticle distribution function, a key quantity that is difficult to find when working with fundamental theories. Goldie and Withington obtained non-linear solutions to the kinetic equations for the coupled quasiparticle and phonon systems~\cite{GoldieWit2012}.
Later, their solutions for the non-thermal quasiparticle distribution function were combined with the Mattis-Bardeen theory when by de Visser et al. proposed a mechanism for microwave suppression on superconducting aluminum resonators~\cite{VisserKla2014}.

Though some of these extensions can be used in the regimes of strong fields~\cite{Gurevich,Kubo}, they conceptualize the notion of $E_1\to E_2=E_1+\hbar\omega$. At large enough fields, however, we shall see that there are quasiparticle surface-states with energies that change dramatically during each cycle of the AC field. The situation we introduce here is reminiscent of the smeared density of states in \cite{Gurevich}, except we include explicitly the depth and time dependence of both the superconducting gap and the quasiparticle states. In particular, we find that $E_2(t)-E_1(t)$ is not constant for such states, which makes the expression $E_2=E_1+\hbar\omega$ unreliable. This distinction demands a new framework to handle such quasiparticle states in strong AC fields. Rather than considering transitions occurring coherently over many cycles in the weak-coupling limit, we take the opposite limit and consider scattering events occurring within a single cycle and ultimately solve the same quantum dissipation problem but starting from the adiabatic limit. 

In Section~\ref{sec:formalism} we introduce the Bogoliubov-de Gennes equations and describe quasiparticle surface states whose energies exhibit strong field dependence. In Section~\ref{sec:mechanism} discuss how to incorporate the quasiparticle states with changing energies into a modified version of the two-fluid model. In Section~\ref{sec:results} we compute the field and frequency dependence of our dissipation mechanism and discuss its relevance to the anti-$Q$ slope. Finally, in Section~\ref{sec:conclusion} we offer concluding remarks and possibilities for further study.



\section{The Bogoliubov-de Gennes self-consistent field method}
\label{sec:formalism}
To find quasiparticle states we solve the Bogoliubov-de Gennes {(BdG)} self-consistent field equations \cite{deGennes}:
\begin{equation}
\begin{aligned}
(H_e+U)u+\Delta v&=Eu\\
-(H_e^*+U)v+\Delta^* u&=Ev
\end{aligned}
\label{eq:BdG}
\end{equation}
with
\begin{equation*}
H_e(\mathbf{r})=(-i\hbar\nabla-e\mathbf{A}(\mathbf{r}))^2/2m-E_\textrm F,
\end{equation*}
where $e$ is the fundamental charge, $m$ is the mass of an electron, and $E_\textrm F$ is the Fermi energy. These equations must be solved self-consistently with the potentials given by
\begin{equation}
\begin{aligned}
U&=-V\sum_n |u_n|^2 f_n + |v_n|^2 (1-f_n)\\
\Delta&=V\sum_n u_nv_n^*(1-2f_n),
\label{eq:potentials}
\end{aligned}
\end{equation}
where $f_n$ is the occupancy of state $n$ and $V$ is a constant describing the strength of the interaction. The expression for $U$ corresponds to a Hartree-Fock potential that scales with the electronic density. To avoid simulating a semi-infinite half-space, we limit the domain to a depth $L$ and employ a hard-wall potential at $z=0$. Provided $L\gg\lambda$, this truncation does not affect the energies of states localized near the surface. Solving these expressions numerically begins with an initial guess for the potentials, solving for $(u_n,v_n)$ and subsequently refining the estimated potentials, and iterating until the process settles on consistent solutions for $(u_n,v_n)$, $U$, and $\Delta$. Note that Eqns.~\eqref{eq:BdG} possess a symmetry: If $(u,v)$ is a solution with energy $E$, $(v^*,-u^*)$ is a solution with energy $-E$. Taking advantage of this symmetry requires also replacing the state's occupation $f$ with $1-f$ to preserve Eqns.~\eqref{eq:potentials}. In that case, some quasiparticles states as treated as `quasi-holes' with opposite energy and occupation.

We solve Eqns.~\eqref{eq:BdG} at the surface of a superconductor in a parallel magnetic field: $\mathbf{A}=A_0\sin(\omega t)e^{-z/\lambda}\hat{y}$, where $\lambda$ gives the London penetration depth of the superconductor and $\hat{z}$ is parallel to the surface normal. We treat $\lambda$ as constant because $\lambda$ does not change significantly for RF fields below $H_{c1}$ \cite{Carlson}. We choose to impose the vector potential externally, rather than using the current density and Maxwell's equations to solve for a new vector potential after each iteration of the self-consistency process, but we confirm that our current density is consistent with our vector potential after reaching self-consistency with $U$ and $\Delta$.

Factoring out the dependence in the directions parallel to the surface, $u(\mathbf{r})\to e^{i(k_xx+k_yy)}u(z)$ and likewise for $v$, we obtain a coupled system
of differential equations: 
\begin{widetext}
\begin{equation*}
\begin{aligned}
\left\{\frac{\hbar^2}{2m}\left[k_x^2+\left(k_y-\frac{eA_0}{\hbar c}\sin(\omega t)e^{-z/\lambda}\right)^2-\frac{d}{dz}^2\right]-E_F+U(z)\right\}u(z)+\Delta(z) v(z)&=Eu(z),\\
-\left\{\frac{\hbar^2}{2m}\left[k_x^2+\left(k_y+\frac{eA_0}{\hbar c}\sin(\omega t)e^{-z/\lambda}\right)^2-\frac{d}{dz}^2\right]-E_F+U(z)\right\}v(z)+\Delta(z) u(z)&=Ev(z),
\end{aligned}
\label{eq:odes}
\end{equation*}
\end{widetext}
with $k_x$ and $k_y$ as parameters. Since the time dependence of the field $A_0(t)$ is much slower than quantum relaxation times (apart from inelastic scattering), we obtain the time dependence of the eigenstates simply by solving these equations at a series of times.

For boundary conditions we set $u(0)=u(L)=v(0)=v(L)=0$ to confine the quasiparticles to the slab. According to Eqns.~\eqref{eq:potentials}, this condition forces the potentials to vanish at the surface, which may seem to be a problem given that, for instance, $\Delta$ should be a nonzero constant if the field is zero. However, the relevant length-scale for features of $\Delta$ is the correlation length, while the sums in Eqns.~\eqref{eq:potentials} include terms oscillating on the much shorter length scale of $1/k_F$. For niobium, $1/k_F=0.08$ nm \cite{Ashcroft}, about $500$ times smaller {than the correlation length of nearly $40$ nm \cite{Gonnella}}. Accordingly, any oscillations near $\Delta(z=0)$ are only present within a tiny fraction of a penetration depth from the surface and are analagous to Friedel oscillations in electronic densities near a surface \cite{Crommie}. Others who have studied the Bogoliubov-de Gennes self-consistent field method near a surface observe the same phenomenon \cite{Troy}, or a similar phenomenon when this method applied within a tight-binding model framework \cite{Croitoru}. Relatedly, authors studying superconductivity in thin films and wires have also noted oscillations in the energy gap as a function of thickness \cite{blatt1963shape,shanenko2006shape}. It is important to distinguish between oscillations in the energy gap as a function of thickness in a thin film and oscillations in the pair potential as a function of position in a thick slab or half-space, but both effects result from boundary conditions. 

We use parameters realistic for an elliptical SRF cavity made from niobium: a Fermi energy of $E_F=5.32$~eV \cite{Ashcroft}, a bulk gap of $\Delta_0=1.5$~meV, a penetration depth of $\lambda=40$~nm, an operating temperature of $2$~K, a geometry factor of $G=270$~$\Omega$, a frequency of $1.3$~GHz, and accelerating gradients up to $30$~MV/m \cite{Gonnella}. Such gradients correspond to magnetic fields of more than half of niobium's lower critical field $H_{c1}$ at $2$ K \cite{French}.

\section{Dissipation mechanism}
\label{sec:mechanism}
In the absence of a field, there is a uniform pair potential $\Delta(z)=\Delta_0$ and all quasiparticle energies must be larger than $\Delta_0$. Once the field is applied, states localized near the surface with energies less than $\Delta_0$ appear. Additionally, there is a continuum of states de-localized from the surface whose energies change negligibly during the cycle.

Because elastic scattering is quick---DFT calculations of electron--impurity scattering times in niobium suggest an elastic scattering time of $\sim$$10^{-12}$--$10^{-15}$~s, several orders of magnitude quicker than typical SRF field frequencies in the GHz range \cite{Liepe}---we can assume that a state stays in equilibrium as long as it can scatter elastically with this reservoir of continuum states. In typical fashion, we take the states having energies above $\Delta_0$ to be in equilibrium, with occupancies determined by the Fermi-dirac distribution $f_n=\left(1+e^{\beta E_n}\right)^{-1}$ for $E_n \geq \Delta_0$.

Bound states with energies below $\Delta_0$ cannot scatter elastically with continuum states and can be filled only by the much slower \textit{in}elastic processes. Because the inelastic scattering rate in niobium of $\sim$$10^{-8}$~s is much slower than $\omega$, the bound states with energies below $\Delta_0$ do not have time to reach equilibrium. Given this relation, along with the divergence of the quasiparticle density of states at $E=\Delta_0$, we take the states having energies below $\Delta_0$ to have occupancies determined by $f_n=\left(1+e^{\beta \Delta_0}\right)^{-1}$ for $E_n < \Delta_0$. Every time the field goes to zero, the bound states return to the continuum and scatter elastically with continuum states. The process repeats when the field starts to increase again and bound states reemerge. 

Instantaneous dissipation is computed using an Ohmic relation 
\begin{equation}
    P_{\text{diss}}(t) = \int dV \Re \left[\frac{1}{\sigma (z,t)}\right] j^2(z,t)
\end{equation}
where $\sigma$ gives the complex conductivity of the material, and the current density is solved from the standard quantum current-density relation,
\begin{equation}
\mathbf j = \frac{1}{2m} \left[ \left(\Psi^* \hat{\mathbf p} \Psi - \Psi \hat{\mathbf p} \Psi^*\right) -2e\mathbf A |\Psi|^2\right].
\end{equation}
The average power dissipated within a given cycle of the RF period is evaluated through
\begin{equation}
\langle P_\textrm{diss}\rangle_\textrm{RF period} = \frac{1}{T}\int_0^T dt \,P_\textrm{diss}(t).
\end{equation}

After computing the dissipation within an RF cycle over an infinitesimal area $ds$, we find the surface resistance $R_\text{s}$ through 
\begin{equation}
\langle P_\textrm{diss}\rangle=\frac12 R_\text{s} \int|H|^2 \, ds
\label{eq:Rs}
\end{equation}
and the quality factor from
\begin{equation}
Q=\frac{\omega U}{\langle P_\textrm{diss}\rangle}=\frac{\omega\mu_0\int |H|^2 \, dv}{R_\text{s}\int |H|^2 \, ds}=\frac{G}{R_\text{s}}
\end{equation}
where $G=\frac{\omega\mu_0\int |H|^2 \, dv}{\int |H|^2 \, ds}$ gives the geometry factor, a parameter independent of the frequency and size of the SRF cavity. For example, to double the frequency of a cylindrical cavity, one must halve the radius of the cavity in order to keep it in the same waveguide mode, so $G$ remains unchanged and truly is a purely geometrical parameter.

\subsection{Elastic scattering of Bogoliubov quasiparticles}
At low temperatures, especially the operating temperature considered here of 2~K, the electronic relaxation time is dominated by elastic scattering with impurities~\cite{Ashcroft,Webb}. To simulate the effects of nitrogen-doped surface treatments in niobium, we calculate impurity scattering rates using a three-dimensional delta function as the perturbing potential to model point-like impurities
\begin{equation}
    \delta V = \alpha \delta(\mathbf r- \mathbf r_0).
    \label{eq:alphadfcn}
\end{equation}
We determine a realistic strength $\alpha$ for the scattering potential from \textit{ab initio} results of DFT calculations incorporating nitrogen impurities in bulk niobium and intuitive formulas derived for a homogeneous electron gas. Specifically, the impurity scattering rate for pure planewave electronic states from this delta-function potential is given by
\begin{equation}
    \tau^{-1}=\frac{2\pi}{\hbar} n_\textrm{imp} \alpha^2 N(0)
\end{equation}
where $N(0)$ denotes the electronic density of states at the Fermi level and $n_\textrm{imp}$ is the concentration of impurities. The Fermi-level density of states is easily deduced from standard DFT calculations and Wannier interpolation techniques provide us with impurity scattering rates in bulk niobium as a function of nitrogen impurity concentration.

With the potential strength deduced, we can then compute elastic scattering rates for the Bogoliubov states we solve for, which involves a scattering rate formula distinct from normal-conducting electrons. Accounting for the broken symmetry at the surface using states of the form $u(\mathbf{r})= A^{-1/2}e^{i(\mathbf k_{||}\cdot \mathbf r_{||})}u(z)$, the impurity scattering rate for Bogoliubov quasiparticles is
\begin{equation}
\begin{split}
    \tau_{n\mathbf k_{||}}^{-1}=&\frac{2\pi L}{\hbar A}\sum_{n'\mathbf k'_{||}}\alpha^2n_\textrm{imp}(z_0)(1-f_{n'\mathbf k'_{||}})\delta(E_{n\mathbf k_{||}} - E_{n'\mathbf k'_{||}})\\
    &\times |u_{n\mathbf k_{||}}(z_0)u_{n'\mathbf k'_{||}}(z_0) + v_{n\mathbf k_{||}}(z_0)v_{n'\mathbf k'_{||}}(z_0)|^2
    \end{split}
\end{equation}
with $L$ and $A$ giving the cavity length and area from the BdG simulation. The distinction from normal electrons is the inclusion of a coherence factor, given here by the factor $|uu'+vv'|^2$, which accounts for the different coherence properties in the superconducting phase.


\begin{figure}
	\centering
	\includegraphics[height=2in]{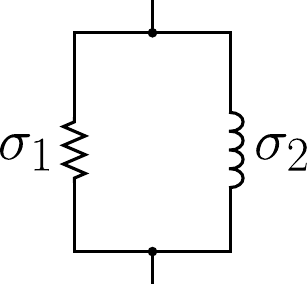}
	\caption[Two-fluid model circuit schematic]{Schematic of the standard text-book definition of the two-fluid model~\cite{Tinkham}. The inductive channel corresponds to the super-fluid consisting of the paired states. The resistive channel corresponds to the normal fluid, consisting of single-particle excitations.}
	\label{fig:2fluid}
\end{figure}
\subsection{Two-fluid model}
\label{sec:2fluid}
The Bogoliubov-de Gennes states we solve for can be used as inputs into the two-fluid model. The standard two-fluid model is depicted in Fig.~\ref{fig:2fluid}, where the two ``fluids'' correspond to the paired states (super-fluid) and the single-particle excitations (normal-fluid) and are modeled as two resistive channels in parallel. The associated conductivities are given by
\begin{equation}
\begin{aligned}
\sigma_1 &= n_n e^2 \tau_n / m \\
\sigma_2 &= n_s e^2 / m \omega
\end{aligned}
\label{eq:conduct}
\end{equation}
where $n_n$ and $n_s$ are the densities of the normal and superconducting electrons, and $\tau_n$ is the relaxation time of the normal electrons~\cite{Tinkham}.  

To estimate the local fraction of normal electrons in the superconducting phase, we compute the proportion of electrons relative to the normal phase available for conduction
\begin{equation}
\frac{n_n}{n}(z)=\frac{1}{N(0)}\sum_i\left( -\frac{\partial f_i}{\partial E_i}\right) \left(u_i^2(z)+v_i^2(z)\right),
\label{eq:frac}
\end{equation}
where $N(0)$ again denotes the electronic density of states at the Fermi level. This quantity, and other normal-phase properties, are readily computed within the Bogoliubov-de Gennes framework by forcing $\Delta=0$ in Eqns.~\eqref{eq:BdG}. For a homogeneous superconductor, Eqn.~\eqref{eq:frac} reduces to the standard BCS form~\cite{Tinkham}. The superfluid density is calculated from the assumption $n_n+n_s=n$.

\begin{figure}
	\centering
	\includegraphics[height=2in]{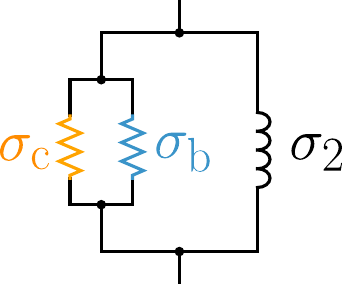}
	\caption[Three-fluid model circuit schematic]{Schematic of the three-fluid model introduced in this work. Like the standard two-fluid model, the inductive channel still corresponds to the superfluid and the resistive channel the normal fluid. However, the normal fluid in the three-fluid model is partitioned into the density arising from bound states with $E_n<\Delta_0$ ($\sigma_b$ in blue) and from continuum states with $E_n\geq\Delta_0$ ($\sigma_c$ in orange).}
	\label{fig:3fluid}
\end{figure}
\subsection{Three-fluid model}
\label{sec:3fluid}
We modify the two-fluid model for the case when the bound states have a different relaxation time than the continuum states. In this case, we treat the bound and continuum states as two parallel resistive channels with conductivity
\begin{equation}
\sigma_1 = \left(n_b \tau_b + n_c \tau_c \right) e^2 / m,
\label{eq:3fl}
\end{equation}
where $n_b$ and $n_c$ are the densities of bound and continuum states, and $\tau_b$ and $\tau_c$ are their respective relaxation times. {These densities are calculated using Eqn.~\eqref{eq:frac} but restricting the summation over states with ${E_n<\Delta_0}$ or ${E_n\geq \Delta_0}$.} We refer to this modification the ``three-fluid’’ model (see Fig.~\ref{fig:3fluid}).

\section{Three-fluid model results and discussion}
\label{sec:results}
\begin{figure}
\includegraphics[width=\columnwidth]{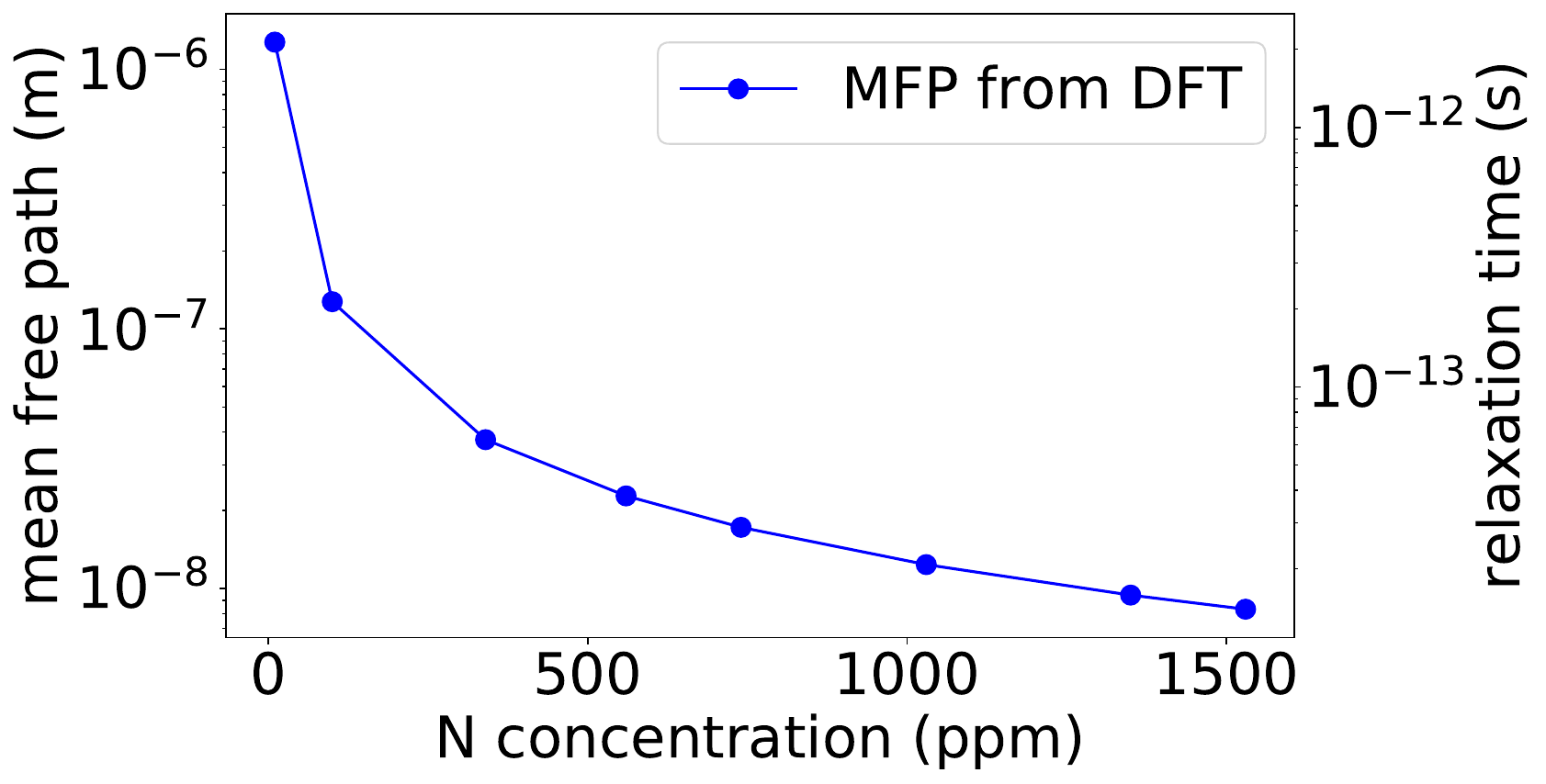}
\caption[Electronic mean free path and elastic relaxation time in niobium vs. concentration of nitrogen impurities calculated with DFT]{Electronic mean free path (left y-axis) and elastic relaxation time (right y-axis) as a function of the nitrogen impurity concentration in parts per million given by DFT calculations of niobium in its normal-conducting phase.}
\label{fig:mfp}
\end{figure}
Figure~\ref{fig:mfp} shows the results of the elastic relaxation time, and associated mean free path, calculated for normal-conducting niobium as a function of concentration of nitrogen impurities.  We find an elastic scattering lifetime of $1.1\times10^{-11}$~s, corresponding to a rate of $9\times10^{10}$~s$^{-1}$, for a nitrogen concentration of one part per million. Considering more realistic impurity concentrations, say $10-10^4$~ppm, we would expect a scattering rate $10^{-12}-10^{-15}$~s. From this result and the calculated Fermi-level density of states, we determine the scattering potential strength of a nitrogen impurity in bulk niobium to be $\alpha\approx 12.13 \,a_0^{3}\,E_\textrm h$. This parameter then enters into the calculation for the elastic scattering rate of Bogoliubov quasiparticles.

\begin{figure}
	\centering
\includegraphics[width=0.5\textwidth]{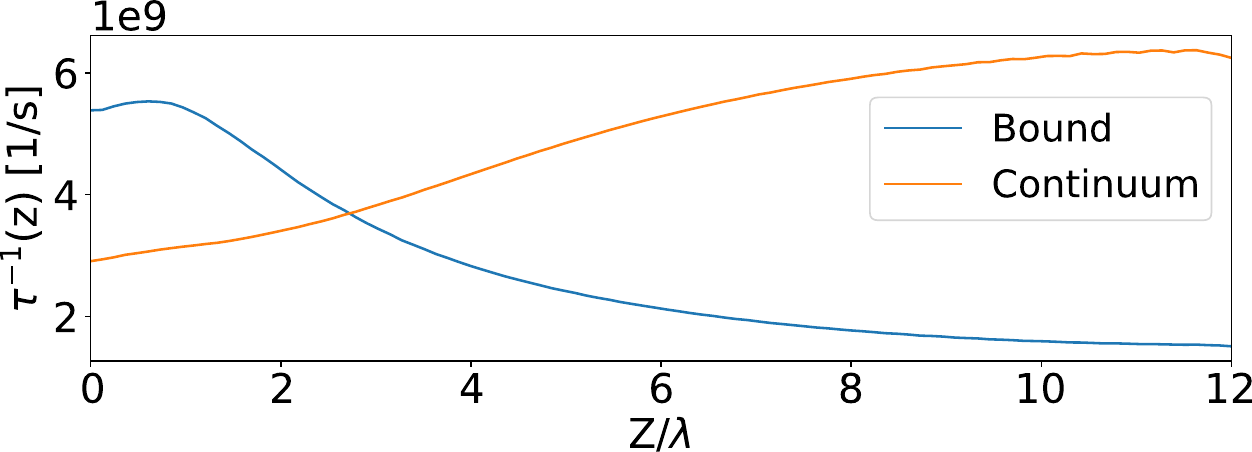}
\caption[Elastic scattering rate of Bogoliubov quasiparticles with point-like impurities vs. depth]{Elastic scattering rate of Bogoliubov quasiparticle at a nitrogen impurity concentration of 1~ppm as a function of depth where the point-like delta-function potential is placed. Partioning the normal fluid
into the two components discussed in Sec.~\ref{sec:3fluid}, the scattering rates for continuum (orange curve) and surface-bound quasiparticle states (blue curve) are considered separately.}
\label{fig:invtau}
\end{figure}

\begin{figure*}
\includegraphics[width=\textwidth]{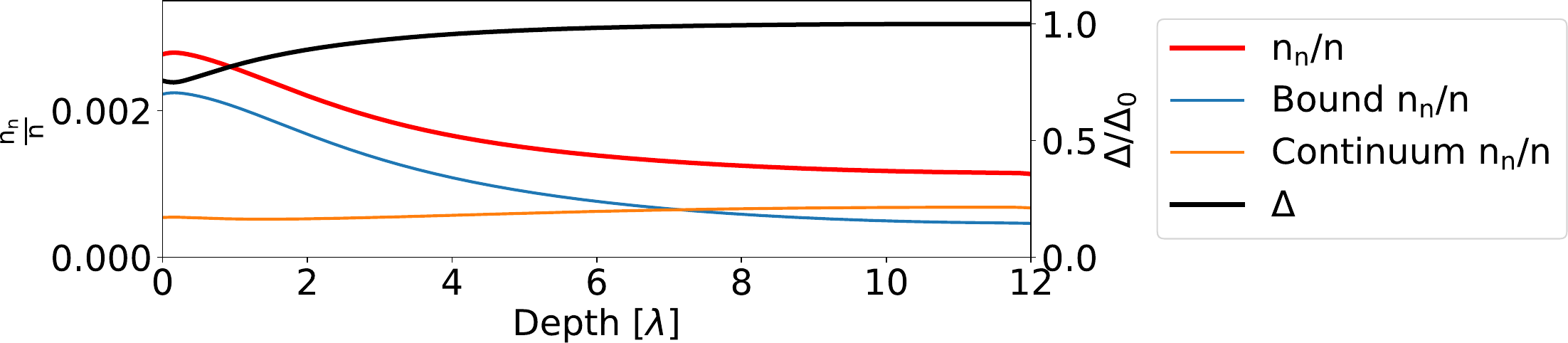}
\caption[Local superconducting gap and normal fluid fraction vs. depth]{Local normal fluid fraction (red curve; left $y$-axis) and local superconducting gap (black curve; right $y$-axis) at $H_0\approx 130$~mT, or $E_\textrm{acc}\approx30$~MV/m. The normal fluid fraction is partitioned into the bound state density (blue curve) and the continuum state density (orange curve), which sums to the total normal fluid fraction $n_n/n$.}
\label{fig:nnfrac}
\end{figure*}

Simulating nitrogen impurities with the perturbing potential given in Eq.~\ref{eq:alphadfcn}, Fig.~\ref{fig:invtau} plots the Bogoliubov quasiparticle elastic scattering rate at a nitrogen impurity concentration of 1~ppm as a function of depth where the delta-function potential is placed. First we consider the effects of the normal-to-superconducting phase transition on the scattering rates. For point-like scatterers away for the surface deep into the bulk of the material, the scattering rate for the bound quasiparticles approaches zero and scattering rate for the continuum states approaches $6\times10^9$~s$^{-1}$ at a nitrogen concentration of 1~ppm. This rate is about a factor of 15 slower than the rate we found for the analogous case in normal-conducting niobium, but a slower scattering rate is expected in the superconducting phase. When a material goes superconducting, a gap opens up at the Fermi level, and quasiparticle scattering rates are expected to scale as
\begin{equation}
\tau^{-1}_\textrm{QP} \sim \tau^{-1}\frac{|\epsilon_k|}{\sqrt{\epsilon_k^2 + \Delta^2}}
\end{equation}
where $\epsilon_k$ is the normal electron's energy measured relative to the Fermi level \cite{Marsiglio1997}. Accordingly, Fermi-level electrons become quasiparticles at the gap edge with infinite lifetimes, but at finite temperatures we need to also consider electrons with energies within $\frac{1}{2}kT$ on either side of the Fermi-level. With niobium's superconducting gap of 1.5~meV, at 2~K the quasiparticle scattering rates can be a factor 17 slower than the normal electrons, agreeing well with the factor of 15 we find here.

We study how the scattering rates for the the bound and continuum Bogoliubov quasiparticle states vary as a function of depth where the delta-function potential is placed. For point-like scatterers near the surface, Fig.~\ref{fig:invtau} shows enhanced scattering rates for the bound quasiparticle states, enhanced by about a factor of two as compared to the continuum states. We also study extended defects by considering a one-dimensional delta function $\delta(z-z_0)$ that can model other defects including surface nanohydrides. For perturbing potentials of this form near the surface, we find scattering rates of the bound quasiparticle states about six times faster than the continuum states. 

As the field turns on, the superconducting gap at the surface weakens caused by the penetrating fields at the surface. Figure~\ref{fig:nnfrac} plots the local superconducting gap (black curve) at an accelerating field of $E_\textrm{acc}\approx30$~MV/m for a simulated cavity 12 penetration depths long. Table~\ref{table:DeltaH} reports the estimated the surface gap reductions {$\Delta({z=0})$} at various values of accelerating fields. 
\begin{table}[ht]
\centering 
\caption[Surface gap reductions estimated from the Bogoliubov-de Gennes self-consistent field method]{Approximate values of the surface gap reduction reported for various field values.} 
\begin{tabular}{c c c c} 
\hline\hline 
$H_0/H_c$ & $H_0$ [mT] & $E_\textrm{acc}$ [MV/m] & $\Delta(z=0)/\Delta_0$ \\ [0.5ex] 
\hline
$H_c/4$ & ~50 & 12 & 0.97\\ 
$H_c/2$ &~100 & 24 & 0.9  \\
$2H_c/3$ &~130 & 30 & 0.75-0.8\\
\hline \hline
\end{tabular}
\label{table:DeltaH} 
\end{table}

Also plotted in Fig.~\ref{fig:nnfrac} is the local normal-fluid fraction from Eq.~\ref{eq:frac} (red curve) at an accelerating field of ${E_\textrm{acc}\approx30}$~MV/m. Noting that $df/dE=-f(1-f)/(kT)$, we can solve analytically what BCS theory predicts for $n_n/n$ at $T=2$~K at zero field. We confirm that the  results we predict for $n_n/n$ for zero field and non-zero fields deep into the material both agree with the BCS-predicted fraction within a few percent. 

Partitioning the local normal fluid fraction into the two components discussed in Sec.~\ref{sec:3fluid}, Fig.~\ref{fig:nnfrac} illustrates the portion of the normal fluid arising from bound states with energies $E_n<\Delta_0$ ($n_b/n$; blue curve) and from continuum states with energies $E_n\geq\Delta_0$ ($n_c/n$; orange curve). These two components sum to the total normal fluid fraction, i.e. $n_b+n_c=n_n$. Subsequently, the three-fluid model uses these density profiles along with elastic scattering rates of the two normal fluid components to determine their underlying conductivities from Eq.~\ref{eq:3fl}, and eventually the cavity's surface resistance from Eq.~\ref{eq:Rs}. Nonetheless, we first decide on optimal simulation parameters through a convergence study of surface resistance in the two-fluid model.

\begin{figure}
	\includegraphics[width=\columnwidth]{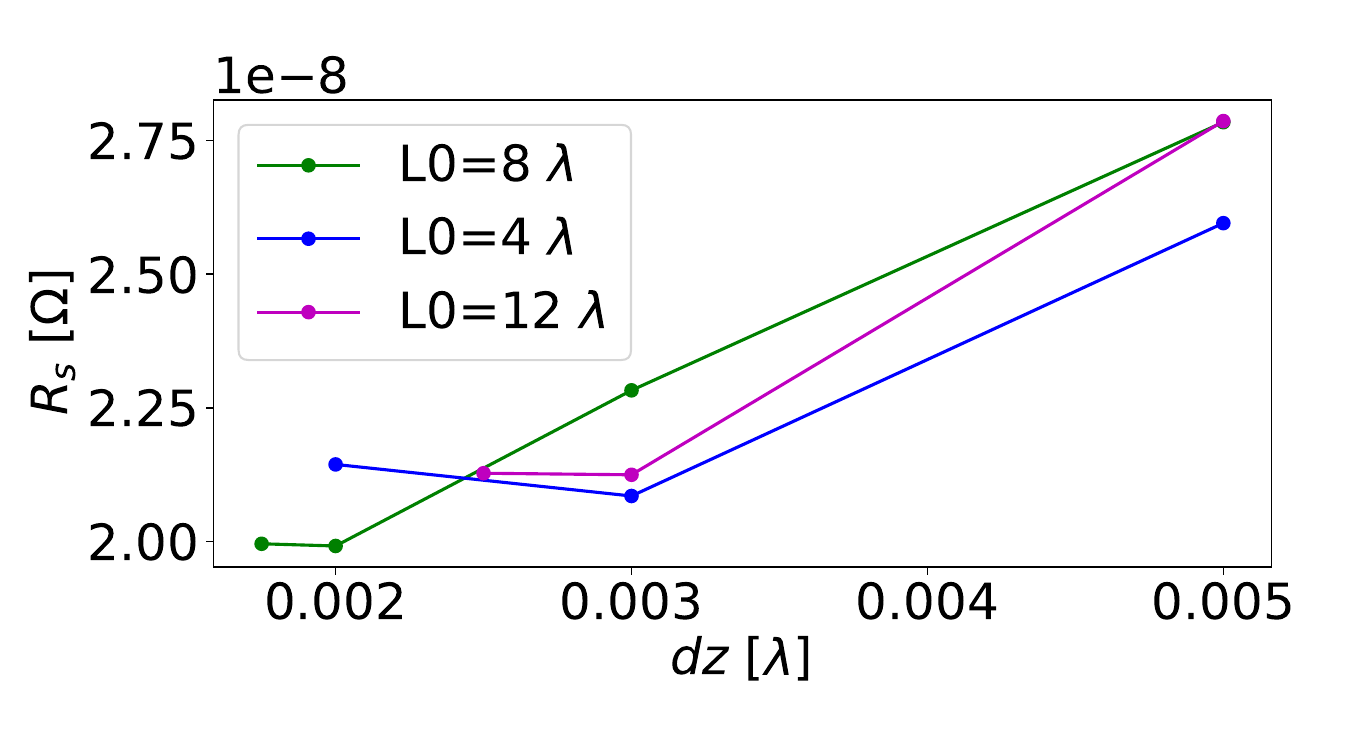}
	\caption[Convergence of surface resistance calculations]{Convergence of calculated surface resistances $R_s$ for various simulated cavity lengths $L_0$ as a function of grid spacing $dz$ in units of penetration depths.}
	\label{fig:Rs_dz}
\end{figure}

\begin{figure*}[t]
	\includegraphics[width=0.95\textwidth]{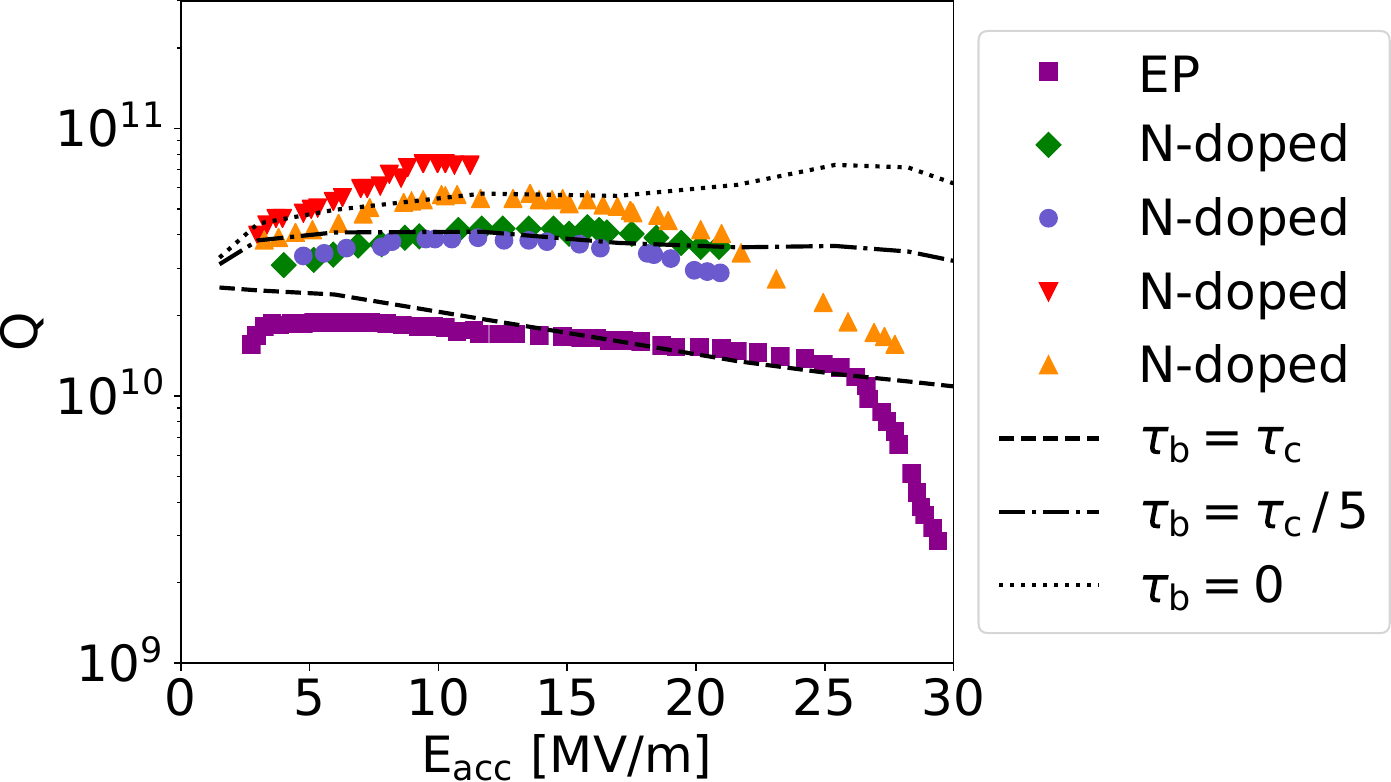}
	\caption[Calculated and measured quality factors vs. field amplitude for a $1.3$ GHz niobium SRF cavity]{Quality factor vs.\ field amplitude for a niobium cavity at frequency $1.3$ GHz. Markers show experimental data from Grassellino et al.\ \cite{Grassellino} for a selection of electro-polished (EP) or nitrogen-doped cavities. The black lines give the results of the three-fluid model for two limiting cases and one intermediate case, all with $\tau_c=0.2$ ps. The dashed lines show the quality factor assuming all states scatter at the same rate, the dotted lines show the quality factor assuming bound states scatter much faster than continuum states, and the dashed-dotted lines show the intermediate case of bound states scattering about a factor of 5 faster than continuum states.}
	\label{fig:QvsB}
\end{figure*}

The three-fluid model collapses to the two-fluid model when the elastic scattering rates of the bound and continuum states are the same, when $\tau_c=\tau_b$. Figure~\ref{fig:Rs_dz} shows the convergence of the surface resistance from the two-fluid model with respect to the grid spacing $dz$ and total cavity length $L_0$, at an accelerating field of $\sim$$30$~MV/m and an elastic scattering time of 0.02~ps. The most sensitive simulation parameter is the grid spacing $dz$, which requires a Nyquist frequency small enough to resolve states which rapidly oscillate with periods near $\sim$$1/k_\textrm F$. For models assuming a spherical Fermi surface, this condition requires 
\begin{equation}
    \frac{\pi}{dz}>2 k_F\Longrightarrow dz<0.0033\, \lambda.
    \label{eq:dz}
\end{equation} 
In accordance with Eq.~\ref{eq:dz}, Fig.~\ref{fig:Rs_dz} shows that the surface resistance is adequately converged at $dz=0.003\,\lambda$ for wide range of cavity lengths. Because we are mostly interested in changes in the logarithm of the quality factor, convergence within a factor of two is sufficient. The $k$-grid spacing is a much less sensitive parameter, and increasing the resolution in $k$-space by a factor of two, from $\Delta k =10\pi$ to $\Delta k =5\pi$, changes $R_s$ by $<0.5\%$. Our upcoming predictions for quality factors as a function of field are computed from simulations employing a $k$-grid spacing of $\Delta k =10\pi$, a $z$-grid spacing of $dz=0.003\,\lambda$, and a total cavity length $L_0=4\,\lambda$.

We compute the dissipation according to the three-fluid model using the density profiles of the bound and continuum normal-fluids to determine the underlying conductivities in Eq.~\ref{eq:conduct}. The elastic scattering times, $\tau_c$ and $\tau_b$, will depend on the nature of the scattering, and fully accurate descriptions of surface scattering include dependencies on additional factors such as impurity concentration profiles, surface roughness~\cite{Jacob} or coatings which include the Nb$_3$Sn coating that is often applied to the surface of SRF cavities~\cite{CarlsonPos2021}. However, the ratio of the timescales for $\tau_c$ and $\tau_b$ is the factor determining how the quality factor changes across field amplitudes.

Figure~\ref{fig:QvsB} plots the results for the quality factors we estimate vs.\ field amplitude, along with several experimental measurements from Grassellino et al.~\cite{Grassellino} for comparison. Given the uncertainty in the ratio $\tau_b/\tau_c$, Figure~\ref{fig:QvsB} shows the range of plausible $Q$ slopes resulting from the three-fluid model for two limiting cases, $\tau_b=0$ (black dotted line) and $\tau_b=\tau_c$ (black dashed line). The results for $\tau_b=0$ roughly align with the anti-$Q$ slope in the experimental data for the first $15$~MV/m, whereas the $\tau_b=\tau_c$ curve is similar to that of the electro-polished cavity producing no anti-$Q$ slope. In both these cases, we take $\tau_c$ to be $\sim$0.2~ps, which roughly corresponds to nitrogen impurity concentrations at 750~ppm. 

As the field strength increases, the density of the bound-state fluid near the surface grows while the density of the continuum state fluid shrinks. The increase in bound states exceeds the decrease in continuum states, so if $\tau_b=\tau_c$ the net effect is to increase $\sigma_1$, drawing more current through the resistive channel and dissipating more energy. However, if $\tau_b$ is negligible compared to $\tau_c$, the net effect is to decrease $\sigma_1$, which will result in more current flowing through the inductive channel and ultimately reduce $R_\text{s}$. Likewise, any source of scattering that makes $\tau_b < \tau_c$ can cause a decrease in $\sigma_1$ and eventually produce an anti-$Q$ slope according to our three-fluid model. We estimate an anti-$Q$ slope to appear approximately when $\tau_b<\tau_c/{5}$, given by the dashed-dotted line in Fig.~\ref{fig:QvsB}. {The magnitude of the quality factor could change significantly depending on the value of $\tau_c$, but the difference in the slope of $Q$ comes from the relative scattering rates among the two components. These results suggest that one could manipulate} the $Q$ slope by modulating any scattering sources that affect the bound states more than the continuum states. 

It is worth noting that in the two- and three-fluid models, as in the conventional linear response theory, the dissipation has an $\omega^2$ dependence. If a mechanism with such dependence produces an anti-$Q$ slope, the addition of a mechanism with flat frequency dependence and no anti-$Q$ slope---such as the inelastic scattering mechanism discussed in Ref.~\citenum{Deyo2022}---would explain why the anti-$Q$ slope becomes more pronounced at higher frequencies. While this argument is speculative, it nonetheless demonstrates the potential for bound states to contribute to the anti-$Q$ slope.


\section{Summary}
\label{sec:conclusion}
For superconductors in large AC fields, there are quasiparticle states for which a linear response approach to dissipation is inadequate. We solve the Bogoliubov-de-Gennes self-consistent field equations at the surface of a superconductor in a parallel magnetic field and find bound states with energies less than the bulk superconducting gap to appear at non-zero fields. These states bound to the surface have energies below the value of the bulk superconducting gap, and these energies change throughout an AC cycle. By focusing on these fundamentally non-perturbative bound states, we embark on a stark departure from conventional theories that are derived in the weak-field limit to instead build a theoretical framework from the adiabatic limit. 

By calculating elastic scattering times of Bogoliubov quasiparticles, our estimates suggest that certain kinds of scattering can indeed affect the bound states more than the continuum states. In the case where the relaxation times of the bound states differs from that of the continuum states, we argue that the two-fluid model should be modified into a three-fluid model. The three-fluid model considers two normal fluid in addition to the superconducting fluid, where the continuum and bound quasiparticles are viewed as two separate normal-conducting fluids. The resulting quality factor could either increase or decrease with field strength, depending on whether the relaxation time of the bound state fluid is comparable to or much smaller than that of the continuum state fluid. According to the model we propose, $Q$ slopes can be tuned by controlling the concentration profiles of identified scattering sources in an SRF cavity. Then, tuning the concentration of such scattering sources in an SRF cavity can ultimately activate an anti-$Q$ slope.

Finally, we note that refining these calculations to include \textit{ab initio} electronic band-structures and the material's actual Fermi surface are the sensible next steps to build upon these results. 

\section{Acknowledgments}
This work was supported by the U.S. National Science Foundation under Award PHY-1549132, the Center for Bright Beams.

\bibliographystyle{apsrev4-2}
\bibliography{TUIBA01}

\end{document}